\documentclass[onecolumn,showpacs,floatfix,11pt,nofootinbib]{revtex4}
\usepackage[dvips]{graphics,color}
\usepackage{epsfig}\usepackage{float}
\RequirePackage{amssymb}
\usepackage{makeidx}
\usepackage[brazil]{babel}
\usepackage[latin1]{inputenc}
\usepackage{graphicx}
\usepackage{indentfirst}
\usepackage{color}
\newcommand{\be}{\begin{equation}}
\newcommand{\ee}{\end{equation}}

\newcommand{\ba}{\begin{eqnarray}}
\newcommand{\ea}{\end{eqnarray}}

\begin{document}

\title{Non-$\mathbb{Z}_{2}$ symmetric braneworlds in scalar tensorial gravity}

\author{M. C. B. Abdalla$^{1}$}
\email{mabdalla@ift.unesp.br}
\author{M. E. X. Guimarães$^{2}$}
\email{emilia@if.uff.br}
\author{J. M. Hoff da Silva$^{1}$}
\email{hoff@ift.unesp.br}

\affiliation{1. Instituto de F\'{\i}sica Te\'orica, Universidade
Estadual Paulista, Rua Pamplona 145 01405-900 S\~ao Paulo, SP,
Brazil}

\affiliation{2. Instituto de F\'{\i}sica, Universidade Federal
Fluminense, Av. Gal. Milton Tavares de Souza s/n, 24210-346,
Niter\'oi-RJ, Brazil}

\pacs{04.50.+h; 98.80Cq}

\begin{abstract}
We obtain, via the Gauss-Codazzi formalism, the expression of the
effective Einstein-Brans-Dicke projected equations in a
non-$\mathbb{Z}_{2}$ symmetric braneworld scenario which presents
hybrid compactification. It is shown that the functional form of
such equations resembles the one in the Einstein's case, except by
the fact that they bring extra informations in the context of
exotic compactifications.
\end{abstract}
\maketitle

\section{Introduction}

Advances in the formal structure of string theory point to the
emergence, and necessity, of a scalar-tensorial theory of gravity.
It seems that, at least at high energy scales, the Einstein's
theory is not enough to explain the gravitational phenomena
\cite{novos}. In other words, the existence of a scalar
(gravitational) field acting as a mediator of the gravitational
interaction together with the usual purely rank-2 tensorial field
is, indeed, a natural prediction of unification models as
supergravity, superstrings and M-theory \cite{GSW}. This type of
modified gravitation was first introduced in a different context
in the 60's in order to incorporate the Mach's principle into
relativity \cite{BD}, but nowadays it acquired different sense in
cosmology and gravity theories.

On the other hand, braneworld models have being extensively
studied in the recent years \cite{muitos}. Such models present an
elegant way to solve the hierarchy problem. In particular, in the
Randall-Sundrum model, which can be understood as an effective
compactification of the Horava-Witten model \cite{HW}, the
hierarchy problem is solved due to a warped non-factorable
topology and the set up encodes two mirror 3-branes (domain walls)
embedded in a five-dimensional bulk. The extra dimension,
transverse to the branes, is an $S^{1}/\mathbb{Z}_{2}$ orbifold.

In the context of the Brans-Dicke gravity (the simplest
scalar-tensor theory we have in the literature), two braneworld
models were proposed using a local and a global cosmic strings
(references \cite{AG} and \cite{AG2}, respectively) in order to
establish all the bulk-brane structure. In these models the final
scenario is composed by two extra dimensions, one compactified on
the brane and another transverse to the brane (the so-called
hybrid compactification). From the gravitational point of view, a
consistent way to extract information of codimension one brane
models is using the well-known Gauss-Codazzi formalism. This has
been done in the context of the General Relativity Theory, with
$\mathbb{Z}_{2}$ symmetry \cite{JP} and without such a symmetry
\cite{JP2}. The result obtained in \cite{JP2} is particularly
interesting to treat models which present hybrid compactification.
In the ref. \cite{AG3} the Einstein-Brans-Dicke (EBD) effective
equation on the brane were obtained with $\mathbb{Z}_{2}$
symmetry. Going forward, and motivated by the results of refs.
\cite{AG,AG2,JP2} we turn again our attention to the application
of the Gauss-Codazzi formalism to braneworld models in Brans-Dicke
theory but at this time without the  $\mathbb{Z}_{2}$ symmetry.
The main purpose of this paper is to obtain the effective EBD
equation in this context. This work is organized as follows: In
the next Section we briefly comment the role of the
$\mathbb{Z}_{2}$ symmetry in braneworld models. In the Section III
we obtain the form of the effective EBD equation in a very similar
way of what was done to the Einstein's case \cite{JP2}. And
finally in the Section IV we summarize and conclude our results.

\section{The role of the $\mathbb{Z}_{2}$ symmetry}

For the sake of completeness we expose in brief lines some
important remarks about the role of the $\mathbb{Z}_{2}$ symmetry
in braneworld models. The role of such a symmetry in the
braneworld scenario is multiple. In the extra dimensional
framework this symmetry prevents the existence of off-diagonal
fluctuations on the space-time metric, as in the original
Kaluza-Klein framework. Moreover, the $\mathbb{Z}_{2}$ symmetry
acts as an important element in the quantum effective theory on
the brane; in particular, it is possible to show that with such a
symmetry one obtains the right chirality of standard model
fermions \cite{Sun}. Apart of this, it is well-known that, if
$n_{\alpha}$ is an unitary vector orthogonal to the brane, this
symmetry acts as a signal reversor in the jump of the $n_{\alpha}$
across the brane. Therefore, it determines univocally the
extrinsic curvature in the scope of the so-called Israel-Darmois
matching conditions.

All these properties are desirable in the implementation of the
braneworld set up and provide facility in the execution of several
calculations as well. However, it is not a necessary condition in
the mathematical sense. In one hand, the ``Index Theorems'' seem
to provide a good approach to the chirality problem \cite{EW} and,
in what concerns the extrinsic curvature, a generalization for
this problem has been done in the Gerenal Relativity Theory
\cite{JP2}. In the context of the reference \cite{JP2}, there is a
key piece concerning the application of the Gauss-Codazzi
formalism, which is the mean value of the extrinsic curvature,
$\langle K_{\mu\nu} \rangle$. By definition, the mean value of any
tensorial quantity, say $X$, is given by
\begin{equation}
\langle X \rangle =\frac{1}{2}(X^{+}+X^{-}), \label{1}
\end{equation}
where $X^{\pm}$ are the both limits of $X$ approaching the brane
from both sides $\pm$. It is possible to decompose $K_{\mu\nu}$ in
a traceless and a non-null trace part. The traceless part is
responsible by the shear of the $n_{\alpha}$ vector along  the
brane. The mean value of the shear, encoded in the mean value of
$K_{\mu\nu}$, clearly vanishes if one imposes the $\mathbb{Z}_{2}$
symmetry. However, in an hybrid compactification scenario, where
there is at least one extra dimension compactified on the brane,
the mean value of the shear term is appreciable. For an
illustrative picture one can think about a simply extra dimension
on the brane with $S^{1}$ topology. This is the case found in the
models of refs. \cite{AG,AG2}.

To finalize this Section, we should stress that all the
modifications generated by this symmetry are encoded in the
geometry of the system. The global brane topology remains the
same, independently whether the extra transversal dimension is an
orbifold or not. In other words, this symmetry changes locally the
brane set up (e.g., it includes extra terms in the metric) but the
topological properties remain the same.

\section{Gauss-Codazzi formalism in the non-$\mathbb{Z}_{2}$ symmetric case}

Our goal in this Section is to generalize the Gauss-Codazzi
formalism in a braneworld scenario which presents hybrid
compactification in an arbitrary dimension without
$\mathbb{Z}_{2}$ symmetry in the Brans-Dicke Theory. After some
preliminary considerations we particularize our results for the
models inspired in the refs. \cite{AG,AG2}.

\subsection{Notation and Conventions}

We treat the brane as a submanifold of $(n-1)$-dimensions embedded
in a n-dimensional manifold, the bulk. The one-to-one
correspondence between each point of the brane with the bulk
points enables an induced topology on the brane, with a respective
induced metric. Calling the bulk metric as $g_{\mu\nu}$, the brane
metric is given by $q_{\mu\nu}=g_{\mu\nu}-n_{\mu}n_{\nu}$. Note
that, according to this equation, the brane is a time-like
hypersurface. Defining, in a similar way of the eq. (\ref{1}), the
jump of any tensorial quantity, $[X]$, by
\begin{equation}
[X]=X^{+}-X^{-},\label{2}
\end{equation}
it is easy to see that \be [AB]=\langle A \rangle [B]+[A]\langle B
\rangle ,\label{3}\ee and \be\langle AB \rangle=\langle A \rangle
\langle B \rangle+\frac{1}{4}[A][B]\label{4}.\ee

The Gauss equation, which relates the brane and the bulk Riemann
tensors, is given by (denoting by $\bar{bar}$ quantities on the
brane) \be \bar{R}_{\mu\nu\lambda\alpha}=
R_{\kappa\beta\sigma\rho}q_{\mu}^{\kappa}q_{\nu}^{\beta}q_{\lambda}^{\sigma}q_{\alpha}^{\rho}-2K_{\mu[\alpha}K_{\lambda]\nu},\label{5}\ee
while the Codazzi equation is \be
2\bar{\nabla}_{[\lambda}K_{\nu]\mu}=R_{\alpha\kappa\sigma\beta}n^{\alpha}q_{\mu}^{\kappa}q_{\nu}^{\sigma}q_{\lambda}^{\beta}.\label{6}\ee

The Weyl tensor, $C_{\alpha\beta}^{\;\;\mu\nu}$, is related with
the bulk geometric quantities in an arbitrary dimension by

\be
R_{\alpha\beta}^{\;\;\;\mu\nu}=C_{\alpha\beta}^{\;\;\;\mu\nu}+\frac{4}{n-2}g_{[\alpha}^{\;\;[\mu}R_{\beta]}^{\;\;\nu]}-
\frac{2}{(n-1)(n-2)}g_{[\alpha}^{\;\;[\mu}g_{\beta]}^{\;\;\nu]}R\label{7}.\ee

The equations (\ref{5}), (\ref{6}) and (\ref{7}) together give \be
\bar{R}_{\mu\nu}=Y_{\mu\nu}+KK_{\mu\nu}-K^{\lambda}_{\mu}K_{\lambda\nu}\label{8},\ee
where the

\be Y_{\mu\nu}=
\frac{n-3}{n-2}R_{\alpha\beta}q_{\mu}^{\alpha}q_{\nu}^{\beta}+\frac{1}{n-2}R_{\alpha\beta}q^{\alpha\beta}q_{\mu\nu}-
\frac{1}{n-1}Rq_{\mu\nu}+E_{\mu\nu}\label{9}, \ee and
$E_{\mu\nu}=C_{\alpha\gamma\beta\delta}n^{\gamma}n^{\delta}q_{\mu}^{\alpha}q_{\nu}^{\beta}$.
It is important to remark that the extrinsic curvature, as well
the $Y_{\mu\nu}$ tensor belong to the hypersurface
($Y_{\mu\nu}n^{\nu}=0=K_{\mu\nu}n^{\nu}$). Another important
characteristic of the $Y_{\mu\nu}$ tensor is that the trace $Y$ is
related with the Einstein tensor, $G_{\mu\nu}$, by
$Y=-2G_{\mu\nu}n^{\mu}n^{\nu}$. Certainly, working in the
Brans-Dicke Theory, this term  will bring important contributions
from the dilaton field (\ref{24}).

\subsection{Obtaining the EBD effective equation on the brane}

In order to project the equations on the brane, one needs to apply
the operations defined in (\ref{1}) and (\ref{2}) in the eq.
(\ref{8}) taking into account the properties showed in eqs.
(\ref{3}) and (\ref{4}). Therefore, starting with
$[\bar{R}_{\mu\nu}]$, one has\footnote{We should not confuse the
operation defined in eq. (\ref{2}) with the notation used here to
indicate the commutation of indices.}

\be [\bar{R}_{\mu\nu}]=0=[Y_{\mu\nu}]+\langle K \rangle
[K_{\mu\nu}]+[K]\langle K_{\mu\nu} \rangle-\langle
K_{[\mu}^{\;\;\alpha}\rangle[K_{\nu]\alpha}] \label{10}.\ee

The quantities $[K]$ and $[K_{\mu\nu}]$ were already derived  in
the  ref. \cite{AG3} to the case of a six-dimensional bulk with a
five-dimensional brane. We shall particularize the present
analysis to this case from now on. The result to the jump of the
extrinsic curvature was obtained for an empty-bulk model in which
the stress-tensors of the bulk -  $T_{\mu\nu}$ -  and the brane -
$S_{\mu\nu}$ - take the usual form\footnote{It is important to
note that the form of the bulk stress-tensor, with the delta term,
is problematic in a complete cosmological scenario. However, for
the purpose of this paper there is no problem with such a
decomposition.}

\be T_{\mu\nu}=-\Lambda g_{\mu\nu}+\delta S_{\mu\nu}\label{11},\ee
\be S_{\mu\nu}=-\lambda q_{\mu\nu}+\tau_{\mu\nu}\label{12},\ee
where $\Lambda$ is the bulk cosmological constant, $\tau_{\mu\nu}$
is the stress-tensor of the brane regular matter and $\lambda$ is
the brane tension, or the vacuum energy of the brane in the case
of isotropic Poincar\'e invariant branes. With such
decompositions, the jump of the extrinsic curvature in the
Brans-Dicke (BD) gravity is given by (see \cite{AG3} for details)

\be[K_{\mu\nu}]=-\frac{8\pi}{\phi}\Bigg(\tau_{\mu\nu}+\frac{q_{\mu\nu}}{2(3+2w)}((w-1)\lambda-(w+1)\tau)
\Bigg)\label{13},\ee and the trace is \be
[K]=\frac{8\pi(w-1)}{2\phi (3+2w)}(\tau-5\lambda)\label{14},\ee
where $w$ is the BD parameter and $\phi$ is the BD scalar field
(generically called here as ``dilaton").

Substituting the equations (\ref{13}) and (\ref{14}) in the eq.
(\ref{10}) one finds

\begin{eqnarray}
-\Big(\frac{8\pi}{\phi}\Big)^{-1}[Y_{\mu\nu}]&=&\left. \langle
K_{\alpha[\mu}\rangle\tau_{\nu]}^{\alpha}-\Bigg(\tau_{\mu\nu}+\frac{q_{\mu\nu}}{2(3+2w)}((w-1)\lambda-(w+1)\tau)
\Bigg)\langle K \rangle \right. \nonumber \\&+&\left.
\frac{(3(1-w)\lambda-(w+3)\tau)}{2(3+2w)}\langle
K_{\mu\nu}\rangle\right. .\label{15}
\end{eqnarray}
This equation will be particularly important in order to find the
mean value of the extrinsic curvature. For now, let us first
derive a first form to the EBD effective equation. So, applying
the mean value operator in the equation (\ref{8}) we have

\be \langle \bar{R}_{\mu\nu} \rangle=\bar{R}_{\mu\nu}=\langle
Y_{\mu\nu}\rangle+\frac{1}{4}\Big([K][K_{\mu\nu}]-[K_{\mu}^{\;\;\alpha}][K_{\nu\alpha}]
\Big)+\langle K \rangle \langle K_{\mu\nu} \rangle-\langle
K_{\mu}^{\;\;\alpha} \rangle \langle K_{\nu\alpha}
\rangle.\label{16} \ee

With the Ricci tensor on the brane one can contract it with
$q^{\mu\nu}$ to find the scalar curvature. The result reads

\be \bar{R}=\langle Y \rangle
+\frac{1}{4}\Big([K]^{2}-[K^{\mu\nu}][K_{\mu\nu}] \Big)+\langle K
\rangle^{2}-\langle K^{\mu\nu} \rangle \langle K_{\mu\nu} \rangle
\label{17},\ee and finally we can construct the Einstein's tensor
on the brane. In order to appreciate the role of the extrinsic
curvature shear pointed in the Sec. II as well as signals of
deviations arising in the scope of the BD gravity, let us
decompose $Y_{\mu\nu}$ and $K_{\mu\nu}$ as

\be Y_{\mu\nu}=\frac{Y}{5}q_{\mu\nu}+\varpi_{\mu\nu},\label{18}\ee
and

\be K_{\mu\nu}=\frac{K}{5}q_{\mu\nu}+\zeta_{\mu\nu},\label{19}\ee
where
$\varpi_{\mu\nu}=\frac{3}{4}\Big(R_{\alpha\beta}q^{\alpha}_{\mu}q^{\beta}_{\nu}-\frac{1}{5}R_{\alpha\beta}q^{\alpha\beta}q_{\mu\nu}\Big)+E_{\mu\nu}$.
The $\zeta_{\mu\nu}$ term is the responsible by the shear of the
$n_{\mu}$ vector. With the above decompositions the effective
Einstein's tensor on the brane reads

\be
\bar{G}_{\mu\nu}=\bar{R}_{\mu\nu}-\frac{1}{2}q_{\mu\nu}\bar{R}=-\Lambda_{5}q_{\mu\nu}+G_{N5}\tau_{\mu\nu}+\pi_{\mu\nu}+
\langle \varpi_{\mu\nu} \rangle+\frac{3}{5}\langle K \rangle
\langle \zeta_{\mu\nu} \rangle-\langle \zeta_{\mu}^{\;\;\alpha}
\rangle \langle \zeta_{\nu\alpha} \rangle,\label{20} \ee where
$\Lambda_{5}$ is the effective cosmological constant on the
4-brane given by

\be \Lambda_{5}=\frac{3}{10}\langle Y \rangle+\frac{6}{25}\langle
K \rangle^{2}-\frac{1}{2}\langle \zeta^{\alpha\beta} \rangle
\langle \zeta_{\alpha\beta} \rangle+
\frac{3}{8}\Big(\frac{8\pi}{\phi}\Big)^{2}\frac{\lambda(w-1)}{(3+2w)^{2}}(\tau+\lambda(w-1)),\label{21}\ee
the $G_{N5}$ term, which mimics the Newtonian gravitational
constant, is

\be
G_{N5}=\frac{3}{8}\Big(\frac{8\pi}{\phi}\Big)^{2}\frac{\lambda(w-1)}{(3+2w)}\label{22},\ee
and the $\pi_{\mu\nu}$ term, quadratic on the brane matter
stress-tensor, is given by

\begin{eqnarray}
\pi_{\mu\nu}&=&\left.
\Big(\frac{8\pi}{\phi}\Big)^{2}\frac{\tau\tau_{\mu\nu}(w+3)}{8(3+2w)}-\frac{1}{4}\Big(\frac{8\pi}{\phi}\Big)^{2}
\tau_{\mu}^{\;\;\alpha}\tau_{\nu\alpha}+\frac{1}{8}\Big(\frac{8\pi}{\phi}\Big)^{2}\tau^{\alpha\beta}\tau_{\alpha\beta}q_{\mu\nu}
\right. \nonumber\\&-&\left.
\Big(\frac{8\pi}{\phi}\Big)^{2}\frac{\tau^{2}q_{\mu\nu}}{8(3+2w)^{2}}(w^{2}+3w+3)\right.\label{23}.
\end{eqnarray}

It is time to make a pause to analyze some properties of eq.
(\ref{20}). The first term contains the effective cosmological
constant (\ref{21}). As expected for this type of models it may
not be constant. We stress that the $\langle Y \rangle$ term
encodes derivative terms of the scalar field, in fact

\begin{eqnarray}
\langle Y \rangle&=&\left.-2\Bigg(\Bigg\langle
\frac{8\pi}{\phi}T_{\mu\nu}n^{\mu}n^{\nu} \Bigg\rangle
+\Bigg\langle\frac{w}{\phi^{2}}\Big(\phi,_{\mu}\phi,_{\nu}n^{\mu}n^{\nu}-\frac{1}{2}\phi,_{\alpha}\phi,
^{\alpha}\Big) \Bigg\rangle\right.\nonumber\\&+&\left.
\Bigg\langle\frac{1}{\phi}\Big(\phi,_{\mu
;\nu}n^{\mu}n^{\nu}-\frac{8\pi}{3+2w}T\Big)\Bigg\rangle\Bigg),\right.
\label{24}
\end{eqnarray}
where $\phi,_{\mu}=\nabla_{\mu}\phi$. Note that all the angled
bracket terms of the equation (\ref{20}) are zero under
$\mathbb{\mathbb{Z}}_{2}$ symmetry, except $\langle Y \rangle$.
Indeed, when $\mathbb{\mathbb{Z}}_{2}$ is imposed, we recover all
the results found in \cite{AG3}. The second term in (\ref{20})
leads to the effective gravitational constant, which strongly
depends on the brane tension. The factor $\langle \varpi_{\mu\nu}
\rangle$ is a generalization of the Weyl tensor in the
$\mathbb{\mathbb{Z}}_{2}$ symmetric case. The terms containing
$\langle \zeta_{\nu\alpha} \rangle$ arise only due the
non-symmetric scenario. In the light of equation (\ref{19}), it is
the mean of the shear of $n_{\mu}$. It is expected that in a
models of hybrid compactification - with some compact extra
dimension on the brane - the term $\langle \zeta_{\nu\alpha}
\rangle$ is appreciable, since the shear will be certainly
important. Therefore, one see that the imposition of the
$\mathbb{\mathbb{Z}}_{2}$ symmetry in hybrid compactification
models means that some important information is lost in the
projection of the EBD equation. One more remark is important here.
As in the symmetric case, analyzed in the ref. \cite{AG3}, here
there is also a type of coupling between the brane tension and the
$(w-1)$ factor. This two quantities always appear together. It can
led to some cumbersome inconsistence between pure BD gravity
(where the BD parameter is expected to be $\sim 1$) and braneworld
models. We stress that it is not a problem in the context of this
work, since we are dealing with BD theory as an intermediate
sector to develop and formalize braneworld models according to the
tendencies pointed by string theory. Anyway, it is just an
apparent incompatibility. An analysis of consistence conditions,
similar to what was done in ref. \cite{Leblond}, to the BD case
show that there is not a forbidden value to the BD parameter
\cite{AG4}.

It is easy to note from the above equations that all that appears
because of the extrinsic curvature mean value terms. Therefore, in
order to completely solve the system it is necessary to express
$\langle K_{\mu\nu} \rangle$ in terms of useful quantities. We
follow the work done in ref. \cite{JP2} and, since we are
repeating their job\footnote{We refer the reader to the ref.
\cite{JP2} for the details of the calculations in the Einstein's
case. Here, we are just generalizing for the Brans-Dicke case.},
we just give the main necessary steps to isolate $\langle
K_{\mu\nu}\rangle$.

The first part is to find an appropriate redefinition of the brane
stress-tensor in terms of which the isolation of $\langle
K_{\mu\nu}\rangle$ can be done in a simplest way\footnote{Note
that from the equation (\ref{15}) it is not easy to isolate
$\langle K_{\mu\nu}\rangle$.}. We start with a new brane
stress-tensor defined by

\be \hat{\tau}_{\mu\nu}\equiv \tau_{\mu\nu}
+(A\lambda+B\tau)q_{\mu\nu},\label{25} \ee where $A$ and $B$ are
constants to be determined. Now, after  expressing $[K_{\mu\nu}]$
and $[K]$ in terms of $\hat{\tau}$ one can, for $w\neq 1$, find
the constants\footnote{In the case analyzed in ref. \cite{JP2}
those constants are (for a n-dimensional bulk)
$A=-\frac{(n-3)}{2(n-2)}$ and $B=\frac{1}{2(n-2)}$.}
$A=\frac{3(1-w)}{4(3+2w)}$ and $B=-\frac{(w+3)}{4(3+2w)}$ in such
way that the eq. (\ref{10}) gives

\be 0=[Y_{\mu\nu}]+\langle K \rangle
[K_{\mu\nu}]+\frac{8\pi}{\phi} \langle K_{[\mu}^{\;\;\alpha}
\rangle \hat{\tau}_{\nu]\alpha}. \label{26}\ee The quantities are
now expressed in terms of $\hat{\tau}_{\mu\nu}
=\tau_{\mu\nu}+\frac{(3(1-w)\lambda-(w+3)\tau)}{4(3+2w)}q_{\mu\nu}$
by

\be
[K_{\mu\nu}]=-\frac{8\pi}{\phi}\Big(\hat{\tau}_{\mu\nu}-\frac{\hat{\tau}}{3}q_{\mu\nu}\Big)\label{27},
\ee and the trace is \be
[K]=\frac{8\pi}{\phi}\frac{2\hat{\tau}}{5}.\label{28}\ee

Treating $\hat{\tau}_{\mu\nu}$ as a $5\times 5$ symmetric matrix
with $det(\hat{\tau}_{\mu\nu})\neq 0$ we have, by definition,
$(\hat{\tau}^{-1})^{\mu\sigma}\hat{\tau}_{\mu\nu}=\delta^{\sigma}_{\nu}$.
Then, inserting the eq. (\ref{27}) into (\ref{26}) and multiplying
the result by $(\hat{\tau}^{-1})^{\mu\nu}$ one has

\be \frac{8\pi}{\phi}\langle K
\rangle=\frac{3(\hat{\tau}^{-1})^{\mu\nu}
[Y_{\mu\nu}]}{9-(\hat{\tau}^{-1})^{\mu}_{\mu}\hat{\tau}^{\nu}_{\nu}}\label{29}.
\ee To guarantee a finite mean for the trace of the extrinsic
curvature one must impose\footnote{In general, for a n-dimensional
bulk this constraint reads $(n-3)^{2}
\neq(\hat{\tau}^{-1})^{\mu}_{\mu}\hat{\tau}^{\nu}_{\nu}$.} that
$9-(\hat{\tau}^{-1})^{\mu}_{\mu}\hat{\tau}^{\nu}_{\nu} \neq 0$. In
fact, there exist many examples of ordinary matter which
stress-tensor is related with the dimension of the manifold in
question, e. g., topological defects and Chern-Simons term (for
odd dimensions). However, up to our knowledge there is not an
approach which circumvents this problem. Here we assume that
$9\neq (\hat{\tau}^{-1})^{\mu}_{\mu}\hat{\tau}^{\nu}_{\nu}$.
Substituting the eq. (\ref{29}) into (\ref{26}) we arrive at

\be -\frac{8\pi}{\phi} \langle K_{[\mu}^{\;\;\alpha}\rangle
\hat{\tau}_{\nu]\alpha}=
 [Y_{\mu\nu}]+\frac{3(\hat{\tau}^{-1})^{\alpha\beta}[Y_{\alpha\beta}]}{9-(\hat{\tau}^{-1})^{\sigma}_{\sigma}\hat{\tau}^{\gamma}_{\gamma}}
 (-\hat{\tau}_{\mu\nu}+\frac{\hat{\tau}}{3}q_{\mu\nu}),\label{30} \ee
or, in a more compact way,

\be \frac{8\pi}{\phi} \langle K_{[\mu}^{\;\;\alpha}\rangle
\hat{\tau}_{\nu]\alpha}\equiv -[\hat{Y}_{\mu\nu}].\label{nova}\ee

In order to completely isolate the $\langle K_{\mu\nu}\rangle$
term solving the equation above, one has to use the vielbein
decomposition. To do so, it is convenient to work with a complete
basis, say $h_{\mu}^{\;\;(i)}$ $(i=0,1,..,4)$, of orthonormal
vectors constructed by the contraction of an orthonormal matrix
set which represents a local Lorentz transformation and turns
$\hat{\tau}_{\mu\nu}$ (and consequently $\tau_{\mu\nu}$) diagonal.
The orthonormality conditions are given by

\begin{eqnarray}
\left.h^{\mu}_{\;\;(i)}h_{\mu (j)}=\eta_{(i)(j)},\right.\nonumber \\
\sum_{i,j=0}^{4}\eta_{(i)(j)}h_{\mu}^{\;\;(i)}h_{\nu}^{\;\;(j)}=\sum_{j=0}^{4}h_{\mu}^{\;\;(j)}h_{\nu
(j)}=q_{\mu\nu},\label{31}
\end{eqnarray}
where $\eta_{(i)(j)}$ is the Minkowski metric and we do not assume
Einstein's summation convention over the tangent indices $(i)$. In
terms of this frame, the diagonal $\hat{\tau}_{\mu\nu}$ reads

\be
\hat{\tau}_{\mu\nu}=\sum_{i}\hat{\tau}_{(i)}h_{\mu}^{\;\;(i)}h_{\nu
(i)}.\label{32} \ee

Plugging the eq. (\ref{32}) in (\ref{nova}) and contracting the
result with $h^{\mu}_{\;\;(i)}h^{\nu}_{\;\;(j)}$ we arrive, after
some algebra, at

\be \frac{8\pi}{\phi}\langle K_{\mu\nu}\rangle=-\sum_{i,j}
\frac{h_{\mu}^{\;\;(i)}h_{\nu}^{\;\;(j)}}{\hat{\tau}_{(i)}+\hat{\tau}_{(j)}}[\hat{Y}_{(i)(j)}],\label{33}
\ee where $[\hat{Y}_{(i)(j)}]\equiv
h^{\mu}_{\;\;(i)}h^{\nu}_{\;\;(j)} [\hat{Y}_{\mu\nu}]$. The
equation above is the main goal of the vielbein decomposition
application. It allows us to write the mean of the extrinsic
curvature in a suitable isolated way. It is easy to note that the
diagonal term of (\ref{33}) is given by

\be
\sum_{i=j}\frac{h_{\mu}^{\;\;(i)}h_{\nu}^{\;\;(j)}}{\hat{\tau}_{(i)}+\hat{\tau}_{(j)}}[\hat{Y}_{(i)(j)}]
=\frac{1}{2}(\hat{\tau}^{-1})_{\mu}^{\;\;\alpha}[\hat{Y}_{\alpha\nu}],\label{34}
\ee then we can write down the matching condition to the mean
value of extrinsic curvature

\begin{eqnarray}
\frac{8\pi}{\phi}\langle K_{\mu\nu} \rangle
&=&\left.\frac{1}{2}(\hat{\tau}^{-1})_{\mu}^{\;\;\alpha}[Y_{\alpha\nu}]
+\frac{3(\hat{\tau}^{-1})^{\beta\gamma}[Y_{\beta\gamma}]}{2(9-(\hat{\tau}^{-1})^{\sigma}_{\sigma}\hat{\tau}^{\rho}_{\rho})}
\Big(q_{\mu\nu}-\frac{\hat{\tau}^{\rho}_{\rho}(\hat{\tau}^{-1})_{\mu\nu}}{3}\Big)\right.\nonumber\\
&-&\left.\sum_{i\neq
j}\frac{h_{\mu}^{\;\;(i)}h_{\nu}^{\;\;(j)}}{\hat{\tau}_{(i)}+\hat{\tau}_{(j)}}[\varpi_{(i)(j)}],\right.\label{35}
\end{eqnarray}
where, following the standard notation, $[\varpi_{(i)(j)}]\equiv
h^{\mu}_{\;\;(i)}h^{\nu}_{\;\;(j)} [\varpi_{\mu\nu}]$. From the
equation (\ref{35}) one can find the expressions for $\langle K
\rangle$ and $\langle \zeta_{\mu\nu} \rangle$ from (\ref{19}).

In order to complete the analysis, let us write down the effective
EBD projected equation on the brane from eq. (\ref{20}). In the
orthonormal frame the diagonal term reads

\be
\bar{G}_{(i)(i)}=-\Lambda_{5}+G_{N5}\tau_{(i)}+\pi_{(i)}+\langle
\varpi_{(i)(i)} \rangle+\frac{3}{5}\langle K \rangle \langle
\zeta_{(i)(i)} \rangle+\sum_{k}\langle \zeta_{(i)}^{\;\;(k)}
\rangle \langle \zeta_{(i)(k)} \rangle,\label{36} \ee where
$\pi_{(i)}=
\frac{1}{4}\Bigg(\frac{8\pi}{\phi}\Bigg)^{2}\Bigg(\frac{(w+3)}{2(3+2w)}\tau\tau_{(i)}-
\tau_{(i)}^{2}+\frac{1}{2}\Big(\sum_{j}\tau_{(j)}^{2}\Big)-\frac{(w^{2}+3w+3)}{2(3+2w)^{2}}\tau^{2}\Bigg)$.
Moreover, there exist off-diagonal terms in the Einstein's tensor
given by

\be \bar{G}_{(i)(j)}=\langle \varpi_{(i)(j)}
\rangle+\frac{3}{5}\langle K \rangle \langle \zeta_{(i)(j)}
\rangle+\sum_{k}\langle \zeta_{(i)}^{\;\;(k)}\rangle \langle
\zeta_{(j)(k)} \rangle.\label{37} \ee

The new off-diagonal terms in the eq. (\ref{37}) arise in the
non-$\mathbb{Z}_{2}$ symmetric context only. These last two
equations summarize the generalization of the Gauss-Codazzi
formalism to braneworld models without $\mathbb{Z}_{2}$ symmetry
in the Brans-Dicke gravity. We shall comment these results in the
next Section.

\section{Concluding Remarks}

In this work our main goal was to generalize the Gauss-Codazzi
formalism to non-$\mathbb{Z}_{2}$ symmetric braneworlds in the
framework of Brans-Dicke gravity. It is a direct generalization of
our previous work (see, please, ref. \cite{AG3}). The analysis
with such tools are very general and can be applied in a wide
range of models. We must stress, however, that some attention
should be paid to the topological scenario of the model. The
Gauss-Codazzi formalism is quite useful if we are dealing with
codimension one braneworld models, which is in fact the case
analyzed here. Moreover, in order to keep a compact internal space
without orbifolding it, the extra dimension can be endowed with a
$S^{1}$ topology (which is the more natural choice) or a bounded
$\mathbb{R}$ line. In the context of braneworld models, if we
choose the bounded $\mathbb{R}$ line, the cut-offs on the extra
dimension are understood as new branes \cite{AG,AG2,RK}. From a
phenomenological point of view, the multiple brane scenario is
potentially interesting, see for instance ref. \cite{GD}. However,
from the mathematical point of view, the most natural choice would
be the compact $S^1$ space.

The off-diagonal terms of the EBD effective equation on the brane
can extract more information in the case of anisotropic types of
braneworlds. If one studies braneworlds via Gauss-Codazzi
formalism within the scope of $\mathbb{Z}_{2}$ symmetry, the
off-diagonal terms are suppressed of the brane metric. However, in
a hybrid compactification model such off-diagonal fluctuations
naturally arises (the Kaluza-Klein original idea). Besides, in the
cosmological context, off-diagonal terms in the energy-momentum
tensor usually describe anisotropic matter sources. As one can see
from equations (\ref{36}) and (\ref{37}), under the imposition of
$\mathbb{Z}_{2}$ symmetry, the angled brackets become zero and we
lose any information about the anisotropic terms.

The presence of the dilaton field in the effective EBD equation on
the brane can lead to important phenomenological implications. In
particular, models which constrain the dependence of such field to
the transverse dimension are very interesting, since the projected
equations give the possibility of subtle but important deviations
from the General Relativity Theory in the study of many
gravitational systems \cite{Maartens}. In this vein, the
cosmological analysis of such models seems to be a very promising
field of research. More specifically, the study of the scalar
field influence in cosmological systems, targeting a systematic
comparison with braneworld models in General Relativity and with
$\mathbb{Z}_{2}$ symmetry, is quite important in order to give
parameters for future developments in braneworlds. We shall to
refer to those questions in the future. We stress, however, that
the complete influence of the dilaton field is model dependent.

Finally, we do not touch here the problem of stabilizing the
dilaton field because it is out of the scope of this paper. In
principle, it can be obtained by the addiction of a well-behaved
potential in the Brans-Dicke part of the gravitational action
\cite{Perivo}. To finalize, we stress that by the shape of the
dilaton field it is possible to make a fine tuning in the value of
$\Lambda_{5}$ via eqs. (\ref{21}) and (\ref{24}). Of course, it
can also be useful as an {\it ad hoc} argument to stabilize the
distance between the branes in a multi-brane scenario.

\section*{Acknowledgments}

The authors are grateful to the EJPC referee for useful comments
and careful reading of the paper. J. M. Hoff da Silva thanks to
the Departamento de Física/Universidade Federal Fluminense for the
hospitality during the realization of this work and CAPES-Brazil
for financial support. M. C. B. Abdalla and M. E. X. Guimar\~aes
acknowledge CNPq for support.

\end{document}